# Performance evaluation of novel square-bordered position-sensitive silicon detectors with four-corner readout

A. Banu[1,*], Y. Li[1], M. McCleskey[1], M. Bullough[2], S. Walsh[2], C. A. Gagliardi[1], L. Trache[1], R. E. Tribble[1], and C. Wilburn[2]

[1]*Cyclotron Institute, Texas A&M University, College Station, TX 77843-3366, USA*
[2]*Micron Semiconductor Ltd., 1 Royal Buildings, Marlborough Road, Lancing Business Park, Lancing, Sussex, BN15 8SJ, UK*

**Abstract**

We report on a recently developed novel type of large area (62 mm × 62 mm) position sensitive silicon detector with four-corner readout. It consists of a square-shaped ion-implanted resistive anode framed by additional low-resistivity strips with resistances smaller than the anode surface resistance by a factor of 2. The detector position linearity, position resolution, and energy resolution were measured with α-particles and heavy ions. In-beam experimental results reveal a position resolution below 1 mm (FWHM) and a very good non-linearity of less than 1% (rms). The energy resolution determined from $^{228}$Th alpha source measurements is around 2% (FWHM).

## 1. Introduction

There is a continuous need for new types of detectors that provide both position and particle identification to image exotic reactions on the ground and in space experiments. Some detector systems are more complex with an increased granularity and a large number of parameters like in multidetector arrays of modern nuclear physics laboratories, and some are simpler and robust for applications where the number of signals and the algorithms for mass and position reconstruction must be kept to a minimum. The novel position sensitive detectors discussed in this paper fall into the latter category. The detector development and implementation were carried out by a collaboration between Micron Semiconductor Ltd. and the Cyclotron Institute at Texas A&M University (TAMU). The motivation behind the work was triggered by a request from the Japan Aerospace Exploration Agency (JAXA) [1] for large area position sensitive silicon detectors that should fulfill the needs for their current and future scientific space missions.

Among the various JAXA areas of exploration is the Earth's magnetosphere. The magnetosphere is the region of space to which the Earth's magnetic field is confined by the solar wind plasma blowing outwards from the Sun, extending to distances in excess of 60 000 km from Earth. Much has been learned about this dynamic plasma region over the past 50 years since the first direct measurements were made by the early Sputnik and Explorer spacecraft. There exist clues, theories and observations, but only partial understanding. Questions like how is solar wind energy transmitted to the magnetosphere, how are particles injected into the inner magnetosphere during magnetic storms, and what is the origin of trapped particles in the Earth's radiation belts are not answered yet. The observations of both short-term and long-term spatial and temporal evolution of trapped particles are expected to contribute valuable information to the understanding of the transport and loss processes of ions and the origin of these ions in the radiation belt inside the magnetosphere.

---

[*] Corresponding author: Tel: +1-979-845-1411; Fax: +1-979-845-1899
*E-mail address*: banu@comp.tamu.edu (A. Banu)

The ongoing JAXA/NASA joint satellite mission GEOTAIL [2] began in July, 2002, having as a primary objective the study of the structure and dynamics of Earth's magnetotail. This part of the magnetosphere is quite dynamic as it continually absorbs energy from the Sun. The High-Energy Particle (HEP) experiment of the GEOTAIL satellite program was built to explore the aforementioned open questions. The HEP detectors are silicon semiconductor telescopes utilizing the $\Delta E \times E$ algorithm for isotope identification in mass and nuclear charge. The top two layers are position sensitive detectors for determining the particle trajectories as well as energy loss.

A particle telescope with a large geometric factor and with a high mass resolution is required for an isotopic observation of elements from He up to Fe group nuclei in solar energetic particles [3]. An increase in the geometric factor is realized by the use of Si detectors with a large active area, including position-sensitive detectors (PSDs) for particle trajectory determination. The energy deposited in $\Delta E$-detectors varies with the incident angle of a particle to the plane of the detector. To determine the mass of the particles with high resolution, the incident angle must be known so that the uncertainty of the path length is minimized. Hence, when observing energetic ions in space three factors become important in the accuracy of ion identification, namely, the detector position resolution and linearity and the linearity of the energy deposition in the $\Delta E$- and E-detectors.

In the following, we discuss recently developed PSDs in terms of their most important operational parameters: position resolution, position linearity, and energy resolution. The performances of the detectors were evaluated using alpha sources ($^{241}$Am and $^{228}$Th) and various available ion beams like helium, oxygen and copper over a wide range of energies from 7 to 40 MeV/nucleon.

## 2. Description of the novel position-sensitive silicon detector type

### 2.1 Short overview on two-dimensional continuous PSDs

Two-dimensional continuous PSDs can be grouped into two types when the position information is obtained by the use of square-shaped resistive anodes, duo-lateral and tetra-lateral, each type with its own advantages and disadvantages.

In the case of a duo-lateral PSD [4], both sides of the detector are ion-implanted resistive sheets with two parallel electrodes formed at right angles as X and Y contacts, assuring thus a simultaneous measurement of X- and Y-coordinates. The X position signals are extracted from the electrical contacts on the front surface (junction side), while the Y position signals are extracted from the contacts on the rear surface (ohmic side). The advantage of this technique is that distortion-free imaging can be easily obtained in analogy to the one-dimensional case described in [5]. The disadvantage of this type of readout consists in a large leakage current induced by the use of the rear ohmic surface of the detector.

The tetra-lateral PSD type uses only the junction side of the detector for obtaining two-dimensional position having all four signal contacts on the front surface [6]. This kind of PSD structure consists of a reverse biased $p^+$-n junction with extended lateral contacts to the n-layer in a square layout. Compared to the duo-lateral type, the multiple electrodes required on the front surface produce serious distortion in the position pattern — the commonly seen 'barrel' effect [6] — due to interaction between the electrical contacts which occurs near the corners of the active area. But the tetra-lateral type features an easy-to-apply reverse bias voltage and a small dark current.

## 2.2 The circular arc terminated anode: the pin-cushion PSD type

A theoretically distortion-free two-dimensional (2D) image pattern may be obtained using the circular arc terminated anode geometry devised by C. W. Gear [7]. Gear's proposal is shown in Figure 1. The concave borders of radius $a$, represented by black strips, consist of a conductive surface with the resistance per unit length $R_L$, while the middle of the anode structure is a high-resistivity surface with a surface resistance $R_S$. To read out the charges or the corresponding currents of the structure, electrical connections are made at the four corners. As pointed out by Gear, if the sheet resistance $R_S$ of the anode and the resistance per unit length $R_L$ of its border are related by

$$R_L = R_S/a,$$

then the anode exhibits minimal image distortion when the appropriate electronic signal ratio processing is used.

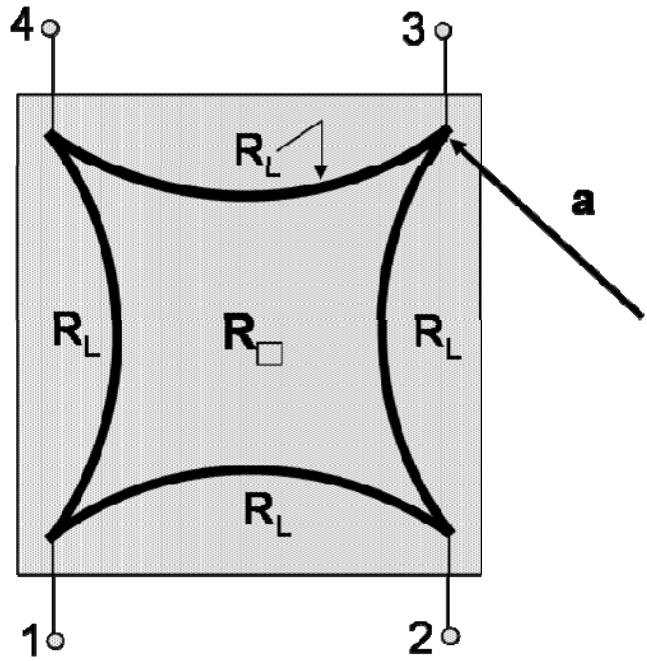

**Figure 1:** Geometry of a circular arc terminated resistive anode. Electrical connections are made at the four corners (see text for details).

Placing the origin of the coordinate system in the middle of the anode surface, a linear relationship can be obtained [8] between the position of the incident particle and the charge or the corresponding current collected at the four corner contacts of the detector:

$$X = \frac{L}{2} \times \frac{(I_2 + I_3) - (I_1 + I_4)}{I_1 + I_2 + I_3 + I_4} \quad \text{and} \quad Y = \frac{L}{2} \times \frac{(I_3 + I_4) - (I_1 + I_2)}{I_1 + I_2 + I_3 + I_4},$$

where $L$ is the distance between two adjacent point-electrodes.

Based on Gear's idea, a nearly distortion-free 2D-resistive anode was developed for a microchannel plate (MCP) electron image system by Augustyniak et al. [9]. The role of the addition of the low-resistance strips in linearizing the 'pin-cushion' distortion commonly seen with four-corner-readout devices is also stressed in [10].

*2.3 Introducing square-bordered PSD type*

The pin-cushion PSD type was first applied to an MCP electron image system and at present is commercially available as position sensitive photodiodes, but the ratio of sensitive area to the total area of the detector is small. The non-sensitive areas result in a considerable reduction of the geometric factor, so PSDs of this type are unsuitable for use in satellite experiments for observing isotope abundances in solar flare/galactic cosmic ray particles, or in general for experiments that require a compact detector system with a large active area. This disadvantage was overcome when Doke et al. [11] developed a new type of pin-cushion PSD with a large active area (62 mm × 62 mm) and good position linearity. The difference between their semiconductive device and what was known before as pin-cushion PSD type [8] was that the geometric configuration of the resistive anode was a *square* and hence $a \to \infty$. Equation (1) shows that for $a \to \infty$, $R_S/R_L$ must also be infinite to keep the two-dimensional image pattern free of distortion. Doke et al.'s anode design concept was to approximate the theoretically desirable value of infinity by large, but practical, values for $R_S/R_L$. They found experimentally that if $R_S$ is larger than $R_L \cdot L$ (where $L$ is the length of the squared-shaped anode) by a factor of 10÷15, the resultant deformation of the position pattern was greatly reduced and the non-linearity was less than 2%. Since then, for two decades their design rule has been used as a reference in the fabrication of this type of large area squared-bordered PSD.

*2.4 The novel square-bordered PSD*

Figure 2 illustrates schematically the squared-bordered PSD type discussed in this paper. The design is similar to previous square-bordered PSDs [11]. However, a much smaller value of $R_S/R_L \cdot L$ is used, leading to less expensive fabrication while obtaining very good performance.

A boron-implanted resistive layer on the n-type Si surface serves as the resistive anode for the charge division in the detector and forms the p-n junction. There is a phosphorus implantation for the n side. The front electrode of the PSD is a square-shaped resistive anode of surface resistance, $R_S$, bordered by additional resistive ion-implanted strips with a low resistance per unit length, $R_L$. The width of the strips, $w$, is 0.5 mm. Four electrical connections, one at each corner of the anode, are formed from aluminium contacts. The four signals generated at these contacts are used to determine the position of incident particles. A fifth signal is taken from the rear side (a non-resistive layer) of the detector via an aluminium-evaporated contact (the common ohmic contact). This signal provides an independent energy measurement with better resolution than can be obtained by summing the four position signals.

**3. Experimental setup**

The characteristics of the newly developed square-bordered position sensitive detectors were investigated in several test experiments using α-particle sources of $^{241}$Am ($E_\alpha$ = 5.47 MeV) and $^{228}$Th (5.34 MeV, 5.43 MeV, 5.69 MeV, 6.06 MeV, 6.23 MeV, 6.78 MeV, 8.79 MeV), and various ion beams — $^{63}$Cu$^{21+}$ at 40, 30, 15 MeV/nucleon, $^{16}$O$^{8+}$ at 60, 45, 30, 16, 7 MeV/nucleon, and $^{4}$He$^{1+}$ at 25, 20, 15, 7 MeV/nucleon — delivered to the Single Event Effects (SEE) beam line of the Radiation Effects Testing Facility [12] from the TAMU K500 Superconducting Cyclotron. A conventional electronic setup for pulse signal processing was employed. Positive voltage was applied to the rear side of the detector through the energy signal preamplifier while the detector front side electrodes were at a ground potential. Each of the four position signals and the energy signal were fed to charge-sensitive preamplifiers. Pulses formed by the preamplifiers were fed further to the main amplifiers (CAEN N568B). The time constant of the pulse shaping was varied as 1μs, 3μs and 6μs. The information on the impinging particle positions and their energy loss in the detector was obtained by pulse height analysis. The pulse heights were digitized by a 16-channel analog-to-digital converter (Phillips Scientific 7164 ADC). The energy signal was also fed to a fast timing filter amplifier (Ortec 474) which generated a timing pulse. This pulse was fed to a constant fraction discriminator and then to a gate generator. The output pulse was used as the gate input of the ADC and to trigger for the data acquisition system. The five digitized signals were recorded for off-line analysis.

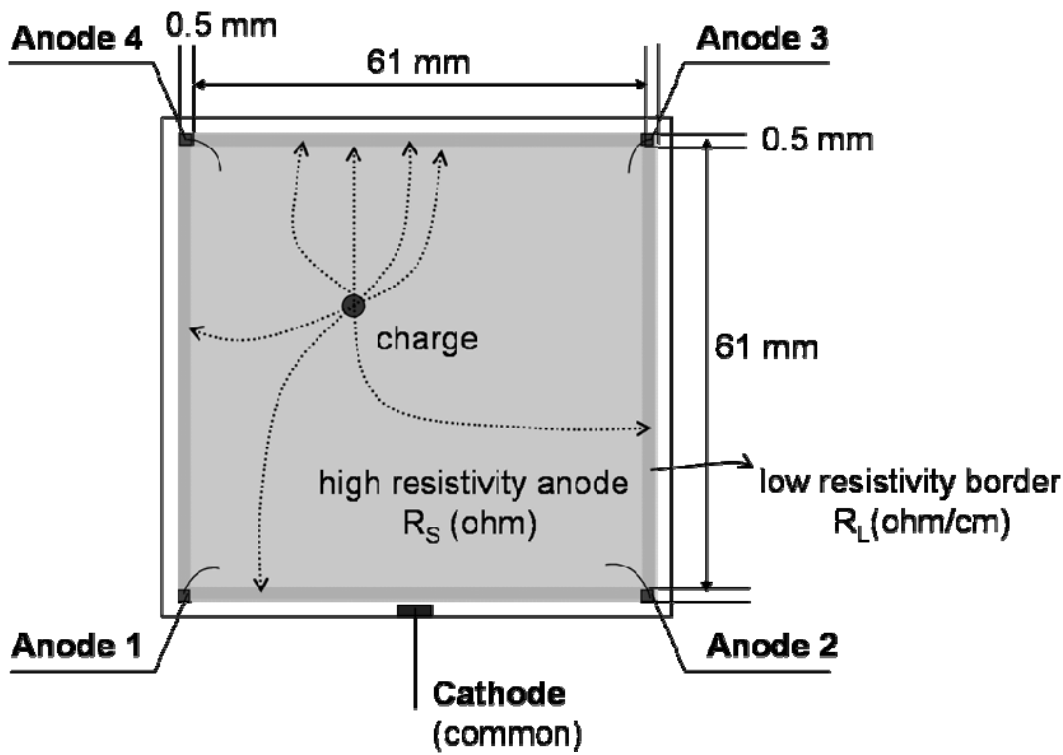

**Figure 2:** A schematic drawing of the two-dimensional position-sensitive silicon detector (PSD) with a resistive anode of surface resistance, $R_S$, bordered by additional ion-implanted strip lines with low resistance, $R_L$, and width of 0.5 mm. The PSD has an active area of 61 mm × 61 mm.

In order to evaluate the main detector characteristics, an Al mask with a regular 11 × 11 grid of pinholes, each with a diameter of 1 mm and spaced at 5 mm, was placed in front of the detectors.

### 4. Detector development studies

In the course of our developmental work more than 25 detectors were investigated. The JAXA requirements were: silicon position-sensitive detectors with an active area of 62 mm × 62 mm, thicknesses of 200 μm, 250 μm, and 400 μm, position resolution better than 1 mm and minimal 2D-image distortion, good energy resolution and low leakage current, to detect and identify ions of He (7-48 MeV/nucleon), Li (8.5-56 MeV/nucleon), C (13-90 MeV/nucleon), O (16-106 MeV/nucleon), and Fe (28-201 MeV/nucleon).

Inspired by the results of previous work by T. Doke et al. [11], we have investigated the role of the low-resistance strips used in the design of square-bordered resistive anodes. Our goal was to determine the ratio of anode surface resistance, $R_S$, to the additional strip resistance, $R_L·L$, that would optimize the detector position linearity and resolution. Table 1 summarizes the characteristics of representative PSDs that were studied.

**Table 1:** Characteristics of some representative position-sensitive silicon detectors investigated.

| Detector Index | Thickness [μm] | Depletion Voltage [V] | Leakage Current [μA] | Anode sheet Resistance [kΩ/ ] | Strip Resistance [kΩ] | Ratio $R_S/R_L·L$ |
|---|---|---|---|---|---|---|
| 1 | 200 | 50 | 1.88 | 40 | 173.6 | 0.23 |
| 2 | 200 | 100 | 0.14 | 20 | 9.92 | 2.01 |
| 3 | 250 | 10 | 1.70 | 2.5 | 31 | 0.08 |
| 4 | 250 | 10 | 0.07 | 2.5 | 18.6 | 0.13 |
| 5 | 250 | 10 | 0.17 | 40 | 173.6 | 0.23 |
| 6 | 250 | 26 | 0.22 | 20 | 9.92 | 2.01 |
| 7 | 400 | 100 | 2.02 | 40 | 173.6 | 0.23 |
| 8 | 400 | 30 | 0.88 | 20 | 9.92 | 2.01 |

Experimental results obtained for 250 μm-thick detectors are presented in Figure 3. Similar results were obtained also in the cases of 200 and 400 μm-thick detectors. The 2D-plot in Figure 3 (a) is from a PSD without the additional bordering resistive strip and, as expected, exhibits a pronounced 'pin-cushion' distortion of the 2D-image pattern. The distortion is less pronounced for the detectors in Figure 3 (b) and (c) and considerably reduced for the detector in Figure 3 (d). These detectors are indexed 3, 4, and 5, respectively, in Table 1. It is important to note that, although in all four cases a collimation Al mask with $11 \times 11$ pinholes was placed in front of the detector, only the detector in Figure 3 (d) exhibits reasonable position resolution, while the resolution is washed out in the other three cases.

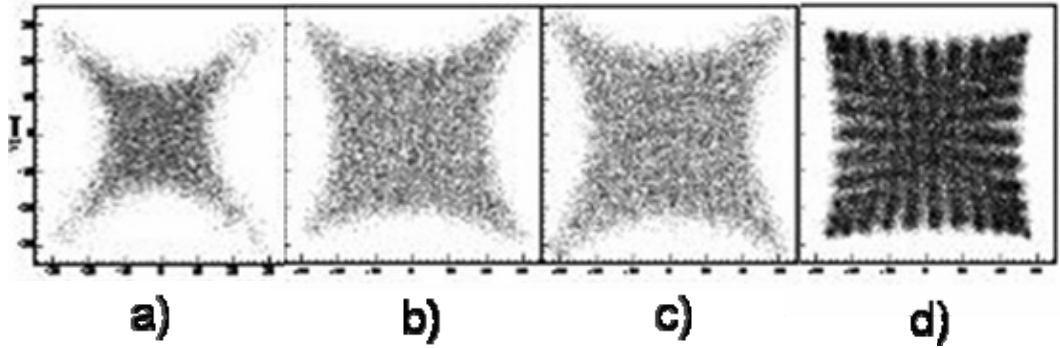

**Figure 3:** Experimental results of α-source measurements on 2D-image patterns for 250 μm-thick PSDs. From left to right the two-dimensional pictures correspond to the detectors with: a) no bordering resistive strip, $R_S$ = 2.5 kΩ/ , b) with aluminium low-resistance strip of 31 kΩ, $R_S$ = 2.5 kΩ/ , c) with low-resistive strip (Si periphery implant) of 18.6 kΩ, $R_S$ = 2.5 kΩ/ , and d) with low-resistance strip (Si periphery implant) of 173.6 kΩ, $R_S$ = 40 kΩ/ , (see text for details).

In the two cases b) and c), the anode surface resistance was $R_S$ = 2.5 kΩ/ and the 2D-plots are alike in terms of 'pin-cushion' distortion pattern, although for the case b) the resistive strip was a thin aluminium film and not a silicon implant as in the other cases. Moreover, the strip resistance had different values — 31 kΩ for b) and 18.6 kΩ for c). The corresponding ratios $R_S/R_L·L$ in the two cases were 0.08 and 0.13, respectively. For the particular case d) which exhibits a much better position resolution and less distortion, the ratio $R_S/R_L \cdot L = 0.23$ and both the anode sheet resistance and strip resistance have much higher values, 40 kΩ/ and 173.6 kΩ, respectively. This may indicate that higher values for anode surface resistance are preferable to achieve both good position resolution and less distortion for the 2D-image pattern.

However, the best results for detector position linearity and resolution were obtained for the ratio $R_S/R_L \cdot L = 2$ corresponding to an anode sheet resistance of 20 kΩ/ and a strip resistance of 9.92 kΩ (see Table 1). This ratio value of 2 is much smaller than the value of 10÷15 found in Doke et al.'s work. The ratio of 2 was a result of taking the value of 20 kΩ/ from previous works [6], [13] as the best starting point for the anode surface resistance, then implanting the periphery resistive strip with the highest ion dose we would consider economically viable. From a manufacturer position, one always looks to minimise the ion implantation dose. Combining this with a strip width small enough so as not to reduce the usable detector area significantly has limited the ratio $R_S/R_L \cdot L$ to a value of around 2. We have thus established a newer design rule in the fabrication of this kind of PSD.

For our research purposes at the Cyclotron Institute/Texas A&M University involving the Momentum Achromat Recoil Spectrometer (MARS) [14], smaller area (20 mm × 20 mm) detectors of this novel PSD type were manufactured following the design rule $R_S/R_L \cdot L = 2$ with $R_S = 20$ kΩ/ and $R_L \cdot L = 10$ kΩ. The borders consist of low-resistivity Si implant strips of length $L = 20$ mm, width $w = 160$ μm, and resistance per unit square $R_L = 80$ Ω/ . Tests similar to the ones done with the JAXA detectors show that these smaller-size detectors give the same distortion-free position response. This makes us conclude that the design rule found works well and is detector-size independent.

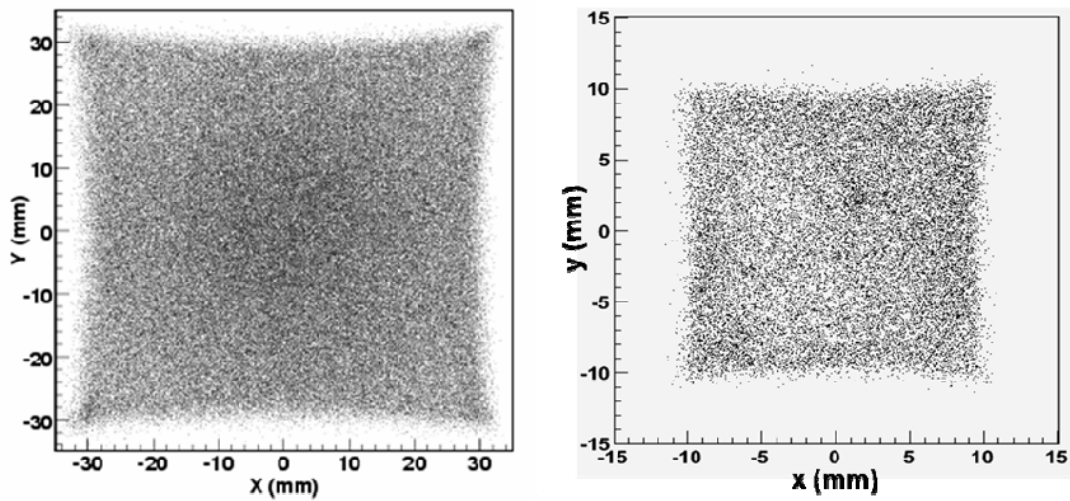

**Figure 4:** Alpha source 2D-image plots (taken without the mask with 11 × 11 pinholes) of a large area (62 mm × 62 mm) 200 μm-thick detector (left) and of a smaller area (20 mm × 20 mm) 1000 μm-thick detector (right), both designed by the rule corresponding to a value of the ratio between the anode sheet resistance and the bounding strip resistance $R_S/R_L \cdot L = 2$.

Figure 4 shows 2D-plots of a large area (62 mm × 62 mm) 200 μm-thick PSD (left) and of a smaller area (20 mm × 20 mm) 1000 μm-thick PSD (right), both manufactured according to the design rule $R_S/R_L \cdot L = 2$. They illustrate nearly distortion-free response for the two-dimensional position information. In the following we present the performance evaluation of these novel square-bordered position-sensitive detectors.

## 5. Results and discussion

For two-dimensional position-sensitive detectors, charge sensitive preamplifier output signals depend on the incident particle position and on the intrinsic time constant, $\tau_{Det} = R\,C_{Det}$, corresponding to the detector characteristics. A pulse-shaping amplifier working with these signals is expected to cause varying ballistic deficits — a reduction of the pulse height depending on the ratio of the amplifier shaping-time constant, $\tau_A$, to the rise time of the detector signal — in position and energy measurements. Significant distortion or non-linearity in the position pattern will result if the ratio $\tau_A/\tau_{Det}$ is small; that is, when the amplifier shaping-time constant is reduced, the charge collection is not completed and a spatial distortion is thus generated. During detector performance evaluations we used CAEN amplifiers with various shaping-time constants of 1 μs, 3 μs and 6 μs for both energy and position signals.

Experimental results are presented next for two categories of detectors (see Section 4, Table 1): in the first category are included detectors corresponding to the ratio $R_S/R_L \cdot L = 0.23$, while in the second category detectors corresponding to the ratio $R_S/R_L \cdot L \cong 2$.

### *5.1 Position resolution and non-linearity*

The PSD signals are contaminated by noise generated by the detector itself and by the subsequent electronics used for signal processing. Therefore to obtain accurate measurements of position and energy, one should consider the properties of both signals and noise. In N. Hasebe et al. [15] results are presented from noise analyses that consider the equivalent circuit of a similar PSD (squared-bordered with four corner readout) and an amplifier system using single RC-differentiation and single RC-integration shaping having equal time constant $\tau(\mu s)$. Taking into account the following noise sources: a) thermal noise of the resistance of the additional line bordering the anode resistive surface layer, b) noise in the FET of the preamplifier, and c) shot noise due to detector leakage current, their mathematical expression for the noise of the position signals $\delta E_n$ in keV (FWHM) for Si material is given by [15]

$$\delta E_n (keV) = \left\{ 12^2 \cdot \frac{\tau \cdot T}{R_{strip}} + (9.8 \cdot C)^2 \cdot \frac{T \cdot R_{eq}}{\tau} + 29^2 \cdot \tau \cdot I \right\}^{1/2}$$

where $T$ is the temperature (K), $R_{strip}$ is the resistance of the strip line connecting the four contacts (kΩ), $R_{eq}$ is equivalent resistance of FET (kΩ), $C$ is the detector capacitance (nF), and $I$ is the leakage current (μA). From here one infers that the noise increases with the square root of the shaping time constant $\tau$ if the thermal or shot noise sources dominate.

The detector position linearity in the *x*-direction (e.g.) has been investigated and results are illustrated throughout the following subsections by corresponding regression analysis figures (see Figures 6 and 8). Data points plotted there are obtained from all 11 *y*-positions corresponding to each of the 11 *x*-positions of the collimating pinhole Al mask. The straight line of the regression analysis in Figures 6 and 8 represents a least squares fit. The root mean square (rms) detector non-linearity (%) for the *x*-position information was determined as

$$\delta_x = \frac{\sqrt{\langle (X_m - X_t)^2 \rangle}}{L},$$

where the indexes *m* and *t* correspond to the measured and true coordinates of each of the mask pinholes, respectively. The rms non-linearity $\delta$ (%) for the whole detector active area is given by:

$$\delta = \frac{\sqrt{\langle(X_m - X_t)^2\rangle + \langle(Y_m - Y_t)^2\rangle}}{L\sqrt{2}} = \sqrt{\frac{(\delta_x^2 + \delta_y^2)}{2}}.$$

*5.1.1 Alpha-particle tests*

Figure 5 shows the dependences of the detector position resolution and linearity on the shaping times used in processing the four-corner signals in the case of a 400-μm thick PSD. For detectors designed with the ratio $R_S/R_L \cdot L \cong 0.23$, best results on position resolution were obtained for an amplification shaping time constant of 3 μs and implied a position resolution of 2.84 ± 0.33 mm (FWHM). In the case of detectors designed with the ratio $R_S/R_L \cdot L \cong 2$, the best position resolution of 2.27 ± 0.31 mm (FWHM) was obtained for an amplification time constant of 1 μs.

With these studies we confirmed the results of previous work [6] and the noise analysis of N. Hasebe et al. [15], that small amplification time constants are necessary to obtain better position resolution for this kind of detector. On the other hand, the detector position linearity does not depend as much on the amplification shaping time constants.

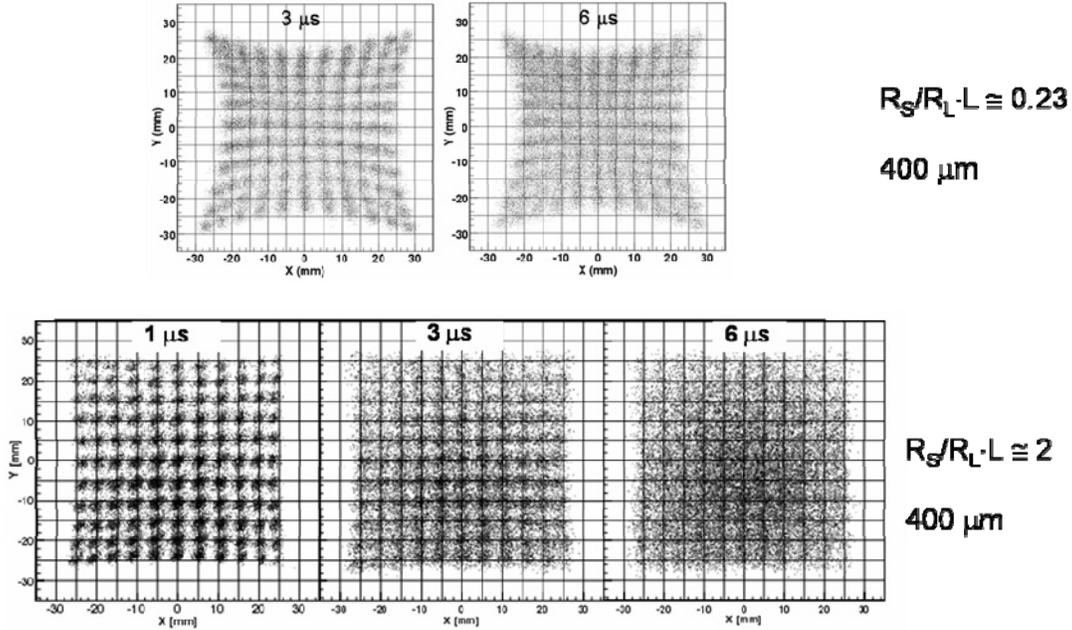

**Figure 5:** Dependence of position resolution and linearity on the shaping times used in processing the PSD corner signals is illustrated by two-dimensional position spectra obtained with the 11 × 11 pinhole mask. These measurements were done with a [241]Am alpha source using 400 μm-thick PSDs and the detector 2D-image plots correspond to the ratios $R_S/R_L \cdot L \cong 0.23$ (top) and $R_S/R_L \cdot L \cong 2$ (bottom).

Example of the detector position linearity $\delta$ (%) (for whole active area) obtained from linear regression analyses are presented in Figure 6 for the case of 400 μm-thick PSDs. For the detectors designed with the ratio $R_S/R_L \cdot L \cong 0.23$, the position distortion is minimal in the detector central area, with a gradual degradation occurring towards the detector edges. The average non-linearity $\delta$ was found to be 2.50 % (rms). A much better response in position linearity, with $\delta$ = 1.15 % (rms), was found for the detectors designed with the ratio $R_S/R_L \cdot L \cong 2$.

*5.1.2 In-beam tests*

For the in-beam tests the energy deposited in the PSDs by heavy ions is much larger than in the case of α-particles, so we expected to obtain considerable improvements in the detector performances. Figure 7 shows an example the results achieved for position resolution (left) and linearity (right) with a 200 μm-thick detector irradiated by an oxygen beam at 7 MeV/nucleon.

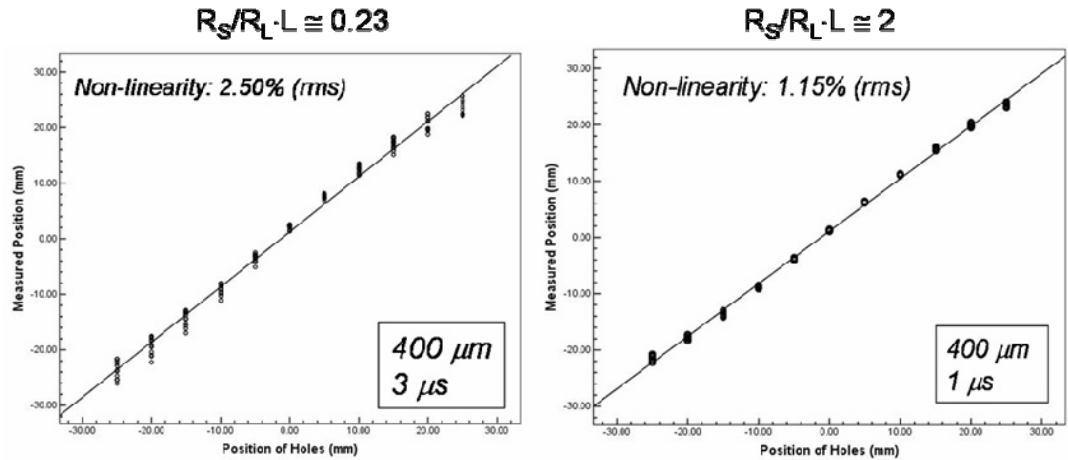

**Figure 6:** Position linearity analysis in the *x*-direction for 400 μm-thick PSDs with $R_S/R_L \cdot L \cong 0.23$ (left) and $R_S/R_L \cdot L \cong 2$ (right). Data points are obtained from the eleven *y*-positions for each of the eleven *x*-positions on the collimation pinhole grid. The straight lines represent the least squares fits for all data points. Non-linearities characterizing the whole detector active area were determined as $\delta = 2.50\%$ (rms) and $\delta = 1.15\%$ (rms) for the two categories of detectors, respectively.

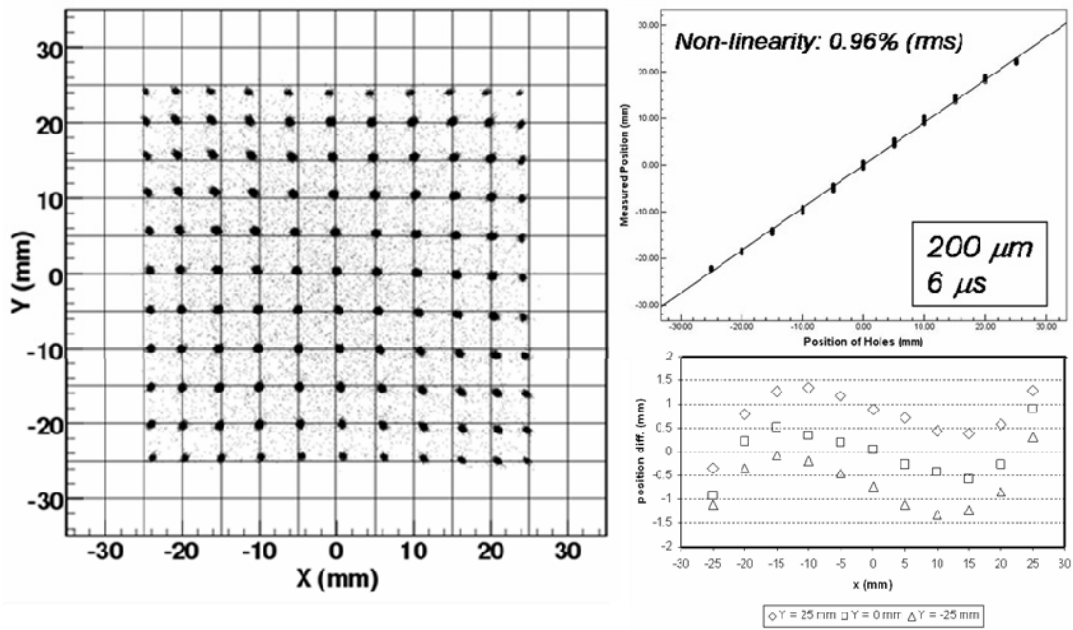

**Figure 7:** In-beam test results on two-dimensional position (left), non-linearity regression analysis (top-right), and the difference between measured and real pinhole positions of the *x*-coordinates corresponding to three y-coordinates — y = +25 mm, y = 0 mm, y = -25 mm (bottom-right), for a 200 μm-thick detector investigated with an oxygen beam at 7 MeV/nucleon.

For the in-beam data the optimum amplifier shaping time constant turned out to be 6 μs. The need for this longer shaping time constant in comparison to the case of α-particle data may be related to additional environmental noise where the in-beam measurements were performed.

A position resolution of 0.73 ± 0.07 mm (FWHM) was found, however, this includes the contribution of the size of the collimation pinholes (diameter = 1 mm). Hence, the detector intrinsic position resolution may be even better. A position non-linearity of 0.96 % (rms) was determined. Though data points obtained from the eleven different y-positions for each x-position are plotted, they are indistinguishable in the figure, reflecting the excellent detector response in position linearity. Similar results were obtained for the He and Cu heavy-ion data.

We find deviations of the *x*-direction measured pinhole positions with respect to the real positions (see Figure 7, bottom-right) to be less than 1.5 mm for the *top* and *bottom* rows on the collimation mask and less than 0.5 mm for the *central* row. Figure 8 illustrates qualitatively the detector position linearity for the *x*-position spectrum corresponding to the central pinhole row on the collimation mask.

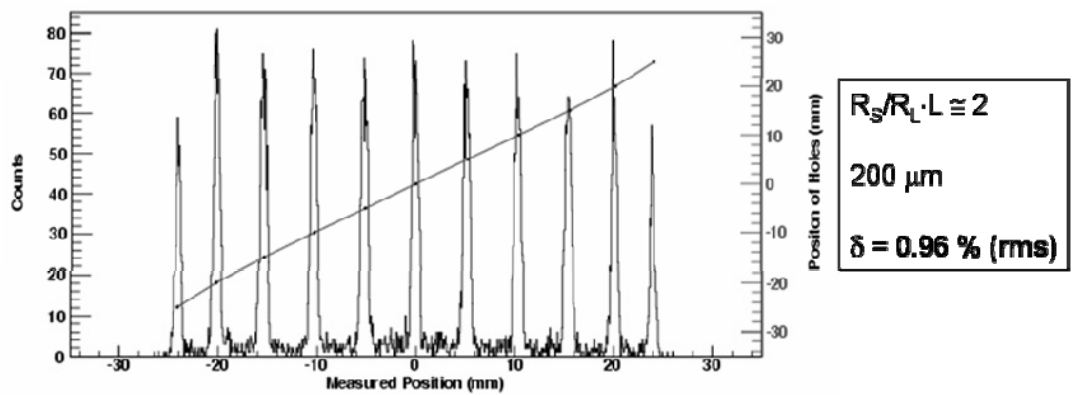

**Figure 8:** Position spectrum in the *x*-direction with linearity curve obtained in the case of a 200 μm-thick detector designed with the ratio $R_S/R_L \cdot L \cong 2$, irradiated by an oxygen beam at 7 MeV/n. The measured position spectrum shows results for particles which pass through the central pinhole row (y = 0 mm) of a 11 × 11 collimation grid. The right-hand axis shows the scale corresponding to the real *y*-positions of the grid pinholes.

### 5.2 Energy resolution

The detector signal from the back electrode was used for energy measurements. Figure 9 illustrates the ballistic deficit for one of the novel 400 μm-thick PSDs designed with the ratio $R_S/L \cdot R_L \cong 2$ and irradiated with α-particles from a $^{241}$Am source.

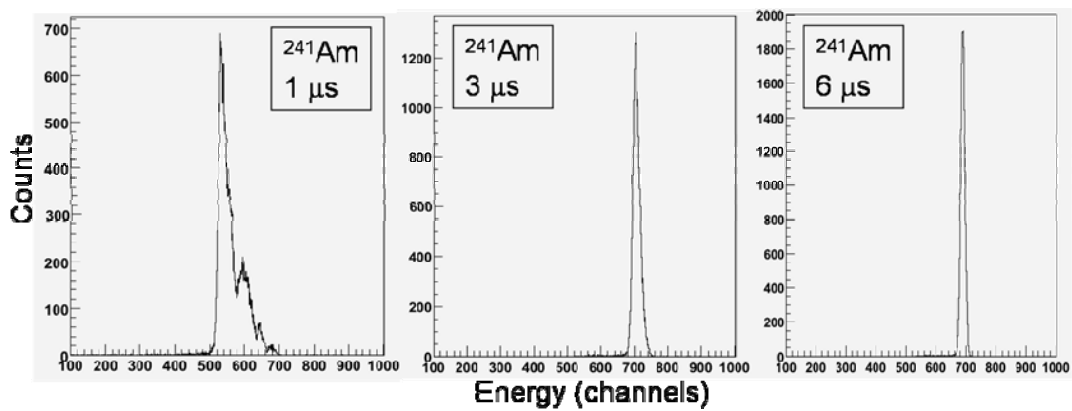

**Figure 9:** Pulse height spectra for 5.5 MeV α-particles incident across the face of a novel 400 μm-thick PSD for various shaping times of the CAEN amplifiers used in the signal processing.

For a 1 μs shaping time constant, the lower peak in the spectrum corresponds to α-particle hits near the detector centre where the ballistic deficit is worst. Events belonging to the higher peak of the distribution are from α-particle hits near the detector corners where the ballistic deficit is minimal even for low shaping times. As the shaping time is increased to 6 μs more charge is collected from the centre of the PSD and the energy distribution becomes Gaussian with a FWHM spread smaller for a 6 μs than for a 3 μs shaping time constant. These results show clearly that amplifiers with longer shaping time constants for processing the energy signal are required to obtain an optimum energy resolution.

For these novel detectors, results on energy resolution are shown in Figure 10 corresponding to data taken with α-particles from a $^{228}$Th source. The energy spectrum on the left is from a 62 mm × 62 mm, 400 μm-thick PSD designed for JAXA and the one on the right is from a 20 mm × 20 mm, 1000 μm-thick PSD for use with MARS. Both detectors have the design ratio $R_S/L \cdot R_L \cong 2$. For α-particles of 6778 keV from the $^{228}$Th source, an energy resolution of 155 keV (FWHM) was determined for the JAXA detector. In this case, the energy spectrum was taken with the collimation Al grid placed in front of the detector. Data taken without the grid demonstrated poorer energy resolution. The MARS detector exhibited much better energy resolution in the order of ~ 68 keV (FWHM) due to its much smaller capacitance. (The area/thickness ratio is smaller by a factor of 24.)

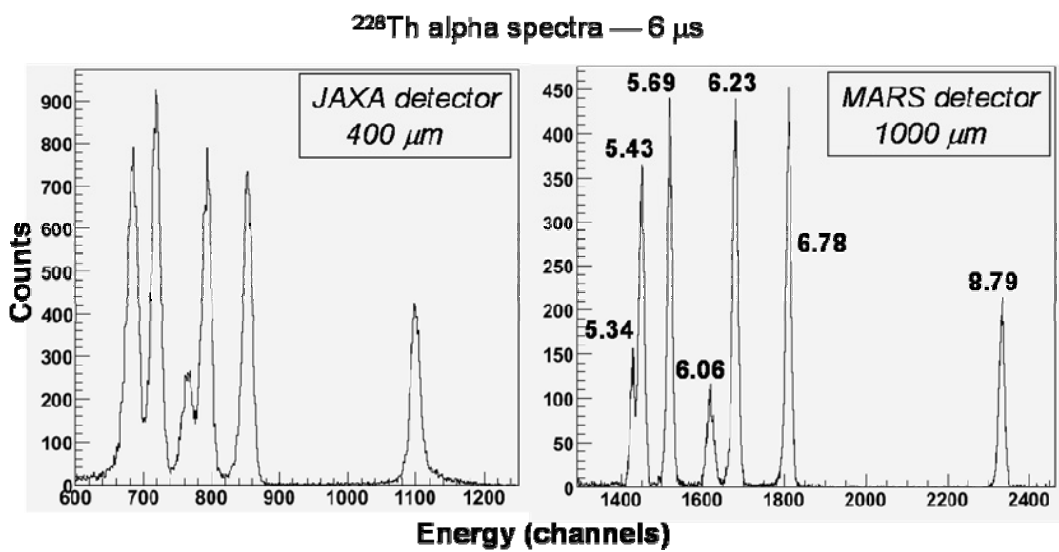

**Figure 10:** $^{228}$Th data illustrating the energy resolution of the novel PSDs. For the JAXA detector the energy spectrum was taken with a collimation mask placed in front, whereas the MARS detector energy spectrum is taken without mask (see text for details).

## 6. Summary

The performance of newly developed two-dimensional square-bordered silicon position-sensitive detectors with four-corner readout of thicknesses 200 μm, 250 μm, and 400 μm have been evaluated using α-particles from $^{241}$Am and $^{228}$Th sources as well as heavy ion beams (He to Cu with energies from 7 to 40 MeV/nucleon) delivered by the TAMU K500 Superconducting Cyclotron. The square-shaped resistive anode of the detectors was bordered by an additional resistive strip line with a resistance smaller by a factor of 2 than the anode resistance. A newer design rule for fabrication of this kind of Si PSD was thus established which differs significantly from previous works.

The characteristics of the novel detectors such as detector position linearity response, position and energy resolutions were investigated. In-beam measurements reveal results for detector position resolution and position non-linearity which were found to be better than 1 mm (FWHM) and less than 1% (rms), respectively. The detector energy resolution was mainly investigated with $^{241}$Am and $^{228}$Th alpha sources yielding a value of around 2% (FWHM).

In terms of PSD signal pulse-shaping, our detector investigation studies confirmed previous experimental results that the position information should be measured by the corner contacts of the front anode electrodes using a short time constant to achieve good position resolution, while the energy information should be measured by the back electrode of the PSD with a longer shaping time constant in order to reduce the ballistic deficit.

These experimental results obtained on the performance of the novel square-bordered position-sensitive silicon detectors show them suitable for use in space-science and nuclear physics experiments.

**Acknowledgements**
The authors gratefully acknowledge the staff of the Accelerator and Radiation Effects Testing Facility divisions at the Cyclotron Institute/Texas A&M University for providing the heavy ion beams and for their assistance in setting up and running the in-beam measurements. This work was supported in part by the U.S. DOE under grant no. DE-FG02-93ER40773 and the Robert A. Welch Foundation under grant no. A-1082.